\documentclass[twocolumn,letter]{jpsj3}

\usepackage{color}
\usepackage{graphicx}% Include figure files
\usepackage{dcolumn}% Align table columns on decimal point
\usepackage{bm}% bold math

\begin{document}

\title{Two qualitatively different superconducting phases under high pressure in single-crystalline CeNiGe$_{3}$}

\author{Shunsaku~Kitagawa$^{1,2}$\thanks{E-mail address: kitagawa.shunsaku.8u@kyoto-u.ac.jp}$^,$\thanks{Present address: Department of Physics, Graduate School of Science, Kyoto University, Kyoto 606-8502, Japan}, Shingo~Araki$^{1,2}$, Tatsuo~C.~Kobayashi$^{1,2}$, and Yoichi~Ikeda$^{3}$}
\inst{$^1$Department of Physics, Okayama University, Okayama 700-8530, Japan \\
$^2$Research Center of New Functional Materials for Energy Production, Storage and Transport,
Okayama University, Okayama 700-8530, Japan \\
$^3$Institute for Materials Research, Tohoku University, Aoba-ku, Sendai 980-8577, Japan
}

\date{\today}

\newcommand{\red}[1]{\textcolor{red}{#1}}

\abst{
We have measured the temperature dependence of resistivity in single-crystalline CeNiGe$_{3}$ under hydrostatic pressure in order to establish the characteristic pressure-temperature phase diagram.
The transition tempearture to AFM-I phase $T_{\rm N1}$ = 5.5~K at ambient pressure initially increases with increasing pressure and has a maximum at $\sim$ 3.0~GPa.
Above 2.3~GPa, a clear zero-resistivity is observed (SC-I phase) and this superconducting (SC) state coexists with AFM-I phase.
The SC-I phase suddenly disappears at 3.7~GPa simultaneously with the appearance of an additional kink anomaly corresponding to the phase transition to AFM-II phase.
The AFM-II phase is continuously suppressed with further increasing pressure and disappears at $\sim$ 6.5~GPa.
In the narrow range near the critical pressure, an SC phase reappears (SC-II phase).
A large initial slope of upper critical field $\mu_0H_{\rm c2}$ and non-Fermi liquid behavior indicate that the SC-II phase is mediated by antiferromagnetic fluctuations.
On the other hand, the robust coexistence of the SC-I phase and AFM-I phase is unusual on the contrary to superconductivity near a quantum critical point on most of heavy-fermion compounds.
}

\abovecaptionskip=-5pt
\belowcaptionskip=-10pt

\maketitle

%\section{Introduction} %% No sections necessary for express letters, letters and short notes
The interplay between superconductivity and magnetism is one of the hottest topics in condensed matter physics. 
It has been believed that superconductivity is mediated by antiferromagnetic (AFM) fluctuations in cuprates and heavy-fermion superconductors since superconductivity is often observed in the vicinity of the AFM quantum critical point at which N\'{e}el temperature $T_{\rm N}$ goes down to zero\cite{T.Moriya_AP_2000,T.Moriya_RPP_2003,P.Monthoux_PRL_1991,C.Pfleiderer_RMP_2009}.
In heavy fermion systems, CePd$_2$Si$_2$\cite{F.M.Grosche_PhysicaB_1996}, CeRh$_2$Si$_2$\cite{R.Movshovich_PRB_1996}, CeIn$_3$\cite{N.D.Mathur_Nature_1998}, and so on, show superconductivity in a narrow pressure range close to a magnetic critical pressure $P_{\rm c}$.
On the other hand, CeCu$_2$$X$$_2$ ($X$ = Si, Ge) and Ce$T$In$_5$ ($T$ = Co, Rh, Ir) exhibit superconductivity in a relatively wide pressure range that is extended toward paramagnetic region\cite{B.Bellarbi_PRB_1984,E.Vargoz_JMMM_1998,R.R.Urbano_PRL_2007,G.Knebel_JPSJ_2008}. 
For CeCu$_2$Si$_2$ and CeCu$_2$Ge$_2$, the superconducting (SC) transition temperature $T_{\rm sc}$ shows a marked increase far from a magnetic critical pressure $P_{\rm c}$, and subsequently disappears at higher pressures. 
The valence crossover scenario was proposed theoretically by K. Miyake and his co-workers to interpret the enhancement of $T_{\rm sc}$ in these materials, where the plausible critical valence fluctuations possibly give rise to the enhancement of $T_{\rm sc}$\cite{K.Miyake_JPSJ_2002,A.T.Holmes_PRB_2004}.

An antiferromagnet CeNiGe$_3$ has an orthorhombic SmNiGe$_3$-type structure with a space group of $Cmmm$ (No. 65, $D_{2h}^{19}$) with inversion symmetry, and also shows pressure-induced superconductivity in a wide pressure range\cite{M.Nakashima_JPCM_2004}. 
An AFM order with the incommensurate propagation vector $k = (0, 0.41, 1/2)$ is observed below $T_{\rm N} = 5.5$~K at ambient pressure\cite{L.Durivault_APA_2002,L.Durivault_JPCM_2003,Y.Ikeda_JPSJ_2015}. 
The pressure-temperature ($P$--$T$) phase diagram was investigated with electrical resistivity measurements on polycrystalline sample\cite{M.Nakashima_JPCM_2004,H.Kotegawa_JPSJ_2006}. 
$T_{\rm N}$ initially increases with increasing pressure up to 3~GPa, and subsequently decreases at higher pressures. 
The critical pressure for $T_{\rm N}$ was estimated to be $P_{\rm c} \sim$  7~GPa. 
Interestingly, partially overlapped two SC domes were observed in CeNiGe$_3$. 
A first SC dome emerges from $\sim$ 1.7~GPa, and is embedded in the deep inside of the AFM phase. 
This SC state becomes fragile at higher pressures, and may disappear in the intermediate pressure range of 4.0-5.4~GPa, where the zero resistance was not observed. 
Subsequently, superconductivity revives in a further high-pressure region, and $T_{\rm sc}$ shows a maximum around the AFM critical point.

The residual resistivity $\rho_0$ strikingly increases with increasing pressure in the previous study: $\rho_0$ at $P_{\rm c}$ is about 100 times as large as that at ambient pressure. 
This striking increase is extrinsic and is possibly caused by an increase in micro-crack and/or residual strain in the polycrystalline sample. 
The large anisotropy of the compressibility in CeNiGe$_3$ was observed in the XRD measurement under high pressure, which was speculated to cause the micro-cracks along the grain boundary\cite{H.Kotegawa_JPSJ_2006}. 
Therefore, the increase in $\rho_0$ indicates a decrease in mean free path of conduction electrons, and may influence the emergence and the broadening of the two SC phases in CeNiGe$_3$.

In this paper, we investigate the unusual SC domes in CeNiGe$_3$ in further detail with electrical resistivity measurement by using a single-crystalline sample under better hydrostatic conditions and clarify the coexistence state of superconductivity with antiferromagnetism in CeNiGe$_3$.

%\section{Experimental}
A single crystal of CeNiGe$_3$ is prepared by the Ni-Ge binary self-flux method following Ref.~\citen{E.D.Mun_PRB_2010}. 
Pressure was applied using two types of pressure cell; an indenter-type pressure cell\cite{T.C.Kobayashi_RSI_2007} for $P \leq$ 4~GPa and an opposed-anvil high-pressure cell designed by K. Kitagawa $et~al.$\cite{K.Kitagawa_JPSJ_2010} for $P >$ 4~GPa. 
We used Daphne 7474\cite{K.Murata_RSI_2008} as the pressure transmitting medium for the indenter cell.
Since Daphne 7474 solidified at 3.7~GPa at room temperature, and the hydrostatic condition becomes worse above 3.7~GPa, we used argon as the pressure transmitting medium for higher pressure experiments using an opposed-anvil high-pressure cell.
From the pressure dependence of $T_{\rm sc}$ in the lead manometer\cite{A.Eiling_JPFMP_1981,B.Bireckoven_JPESI_1988}, we estimated the pressure value as below:
\begin{align}
 P &= \frac{\Delta T_{\rm sc}}{0.364} \hspace{10pt} \text{($P \leq$ 4~GPa)},\notag\\
 P &= \frac{\Delta T_{\rm sc}}{0.364} + (\Delta T_{\rm sc} - 1.456)^{1.6} \hspace{10pt} \text{($P >$ 4~GPa)}\notag
\end{align}
where $P$ is in GPa, and $\Delta T_{\rm sc} = T_{\rm sc}(0) - T_{\rm sc}(P)$.
Low temperature experiments down to 60~mK were carried out by using a dilution refrigerator.
Electrical resistivity was measured with a conventional four terminal method, and the electric current was applied along the $c$-axis.
The results below 4~GPa have already been reported\cite{Y.Ikeda_JPSCP_2014}.

%%%%%%%%%%%%%%%%%%%%%%%%%%% Figure 1 %%%%%%%%%%%%%%%%%%%%%%%%%%%%%%%%%%%%%
\begin{figure}[!tb]
\vspace*{10pt}
\begin{center}
\includegraphics[width=8cm,clip]{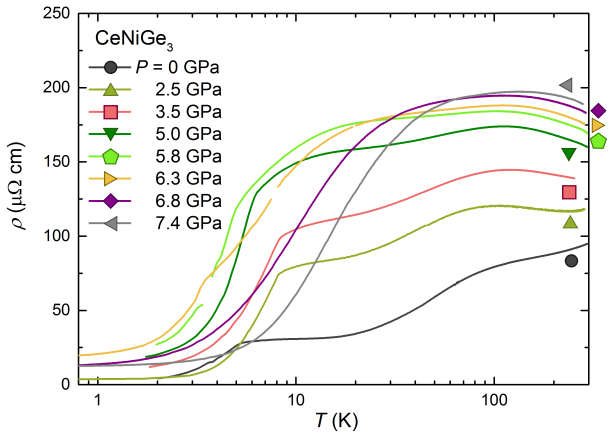}
\end{center}
\caption{(Color online) Temperature dependence of resistivity $\rho(T)$ under high pressures.}
\label{Fig.1}
\end{figure}
%%%%%%%%%%%%%%%%%%%%%%%%%%%%%%%%%%%%%%%%%%%%%%%%%%%%%%%%%%%%%%%%%%%%%%%%%%%

%\section{Results and Discussion}
Figure~\ref{Fig.1} shows the temperature dependence of resistivity $\rho(T)$ measured at several pressures. 
The residual resistivity ratio (RRR) at ambient pressure was evaluated about 30, which is the similar value as that of the polycrystal in the previous study\cite{H.Kotegawa_JPSJ_2006}.
In the low-pressure region, $\rho(T)$ shows a broad bend anomaly resulting from the splitting of the crystal electric field (CEF) at approximately 100~K\cite{E.D.Mun_PRB_2010,A.P.Pikul_PRB_2003}. 
The Kondo effect, whose characteristic temperature was estimated as $T_{\rm K} \sim$ 4.5~K from the specific heat at ambient pressure\cite{A.P.Pikul_PRB_2003}, became significant at high pressures.
The enhancement of $\rho$ with applying pressure can be interpreted as an increase in the $c-f$ hybridization. 
Below $T_{\rm N1}$, the clear reduction of $\rho(T)$ was observed.
Here, $T_{\rm N1}$ was defined as the peak of $- d^2\rho/dT^2$.
In contrast with the previous study on a polycrystalline sample, the striking increase in $\rho_0$ against pressure was not observed for the single-crystalline sample.

%%%%%%%%%%%%%%%%%%%%%%%%%%% Figure 2 %%%%%%%%%%%%%%%%%%%%%%%%%%%%%%%%%%%%%
\begin{figure}[!tb]
\vspace*{10pt}
\begin{center}
\includegraphics[width=8cm,clip]{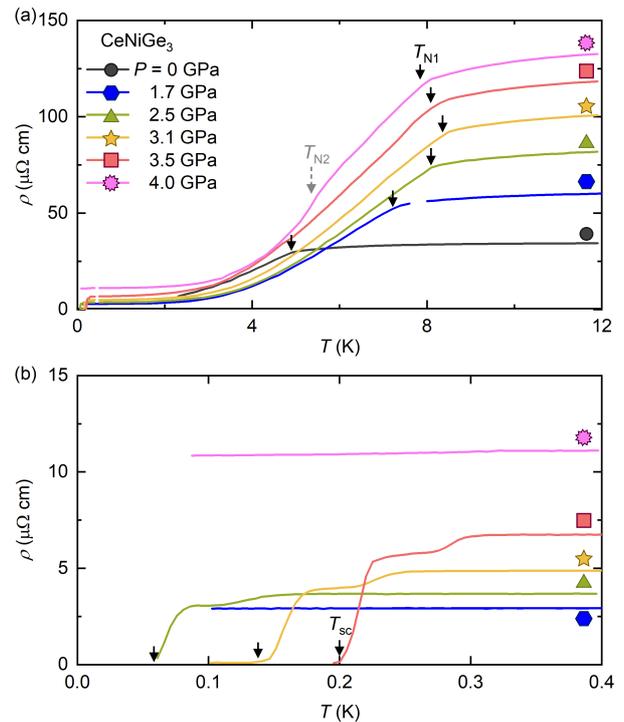}
\end{center}
\caption{(Color online) (a) Temperature dependence of resistivity below 4.0~GPa. 
The black solid (grey dashed) arrows indicate $T_{\rm N1}$ ($T_{\rm N2}$). $T_{\rm N2}$ is observed at 4.0 GPa. 
(b) $\rho(T)$ below 0.4~K. 
The arrows indicate $T_{\rm sc}$.
$T_{\rm sc}$ increases with increasing pressure, and suddenly disappears above 3.5~GPa.}
\label{Fig.2}
\end{figure}
%%%%%%%%%%%%%%%%%%%%%%%%%%%%%%%%%%%%%%%%%%%%%%%%%%%%%%%%%%%%%%%%%%%%%%%%%%%
%%%%%%%%%%%%%%%%%%%%%%%%%%% Figure S %%%%%%%%%%%%%%%%%%%%%%%%%%%%%%%%%%%%%
\begin{figure}[!tb]
\vspace*{10pt}
\begin{center}
\includegraphics[width=8cm,clip]{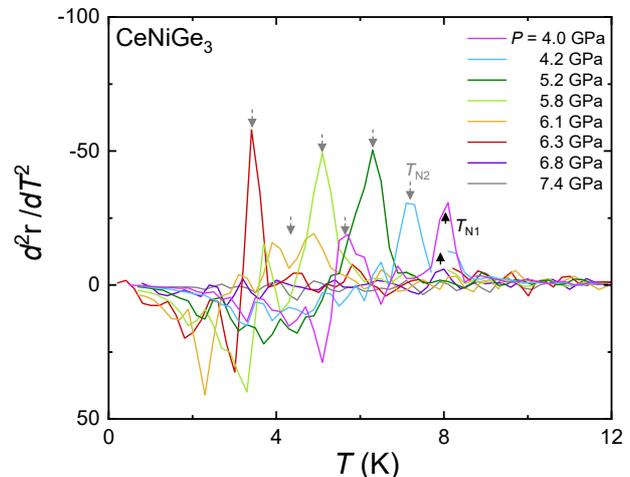}
\end{center}
\caption{(Color online) Temperature dependence of the second derivative of resistivity against temperature}
\label{Fig.S}
\end{figure}
%%%%%%%%%%%%%%%%%%%%%%%%%%%%%%%%%%%%%%%%%%%%%%%%%%%%%%%%%%%%%%%%%%%%%%%%%%%
%%%%%%%%%%%%%%%%%%%%%%%%%%% Figure 3 %%%%%%%%%%%%%%%%%%%%%%%%%%%%%%%%%%%%%
\begin{figure}[!tb]
\vspace*{10pt}
\begin{center}
\includegraphics[width=8cm,clip]{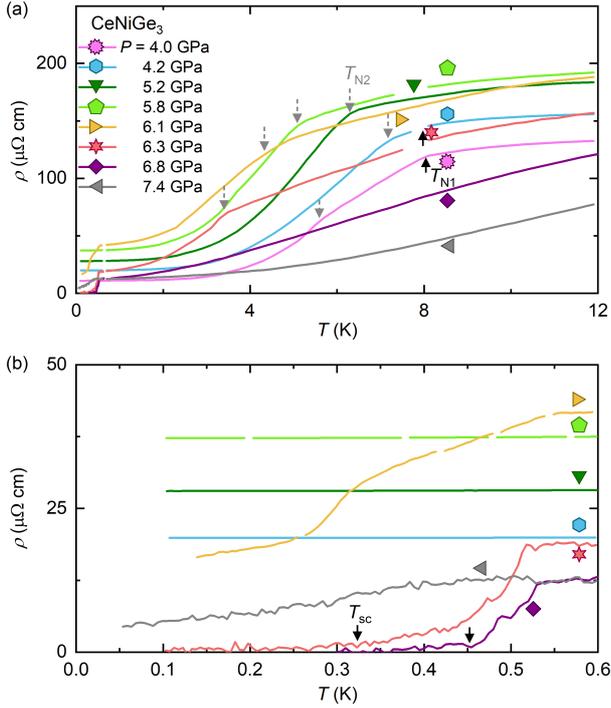}
\end{center}
\caption{(Color online) (a) Temperature dependence of resistivity above 4.0~GPa. 
The black solid (grey dashed) arrows indicate $T_{\rm N1}$ ($T_{N2}$).
(b) $\rho(T)$ below 0.6~K.
The arrows indicate $T_{\rm sc}$.
Zero resistivity is observed at 6.3~GPa and 6.8~GPa.}
\label{Fig.3}
\end{figure}
%%%%%%%%%%%%%%%%%%%%%%%%%%%%%%%%%%%%%%%%%%%%%%%%%%%%%%%%%%%%%%%%%%%%%%%%%%%

First, we focus on the experimental results below 4.0~GPa.
As shown in Fig.~\ref{Fig.2}(a), $T_{\rm N1}$ initially increased with applying pressure and became maximum at 3.1~GPa with $T_{\rm N1}$ = 8.2~K.
As shown in Fig.~\ref{Fig.2}(b), a clear zero-resistivity was observed above 2.5~GPa in this study.
In the previous results\cite{Y.Ikeda_JPSCP_2014}, zero-resistivity was observed at 2.3 GPa, which is consistent with the present study.
The small drop in resistivity before SC transition originates from SC transition in the small part of the sample.
$T_{\rm sc}$ was defined as the temperature at which $\rho$ becomes 0.1$\rho_{0}$.
$T_{\rm sc}$ increased with increasing pressure and suddenly disappeared above 3.5~GPa. 
In addition, the resistivity exhibited an additional kink anomaly at $T_{\rm N2}$ above 3.5~GPa as shown in Fig.~\ref{Fig.2} (a).
This anomaly in $\rho(T)$ can be identified as a clear peak of $- d^2\rho/dT^2$ as seen in the previous experiment\cite{Y.Ikeda_JPSCP_2014}. 
The temperature dependence of $d^2\rho/dT^2$ is shown in Fig.~\ref{Fig.S}.
Both $T_{\rm N1}$ and $T_{\rm N2}$ were suppressed with the application of magnetic field along $a$-axis\cite{Y.Ikeda_JPSCP_2014}, indicating that both magnetic phases can be considered as AFM phases. 
We labeled lower and higher pressure phase AFM-I and AFM-II, respectively.
The apparent SC transition simultaneously disappeared with the emergence of the AFM-II phase. 
These experimental facts demonstrate that the ordered state in the AFM-II phase is conclusively different from the AFM-I state.

With further increasing pressure, a narrow SC region appeared around the AFM-II quantum critical point $P \sim$ 6.5~GPa.
Figure~\ref{Fig.3} shows the temperature dependence of $\rho(T)$ above 4.0~GPa.
Above 3.1~GPa, $T_{\rm N1}$ gradually decreased with increasing pressure.
On the other hand, $T_{\rm N2}$ strikingly increased with increasing pressure and two transition temperatures merge at 4.4~GPa.
After merging, $T_{\rm N2}$ goes to zero and disappears and the pressure dependence of $\rho_{0}$ shows a peak at around 6.5~GPa.
Zero resistivity was observed close to $P_{\rm c}~\sim$ 6.5~GPa and quickly disappeared with further increasing pressure.
$T_{\rm sc}$ was defined in the same manner as that in the SC state below 3.5~GPa.

%%%%%%%%%%%%%%%%%%%%%%%%%%% Figure 4 %%%%%%%%%%%%%%%%%%%%%%%%%%%%%%%%%%%%%
\begin{figure}[!tb]
\vspace*{10pt}
\begin{center}
\includegraphics[width=8cm,clip]{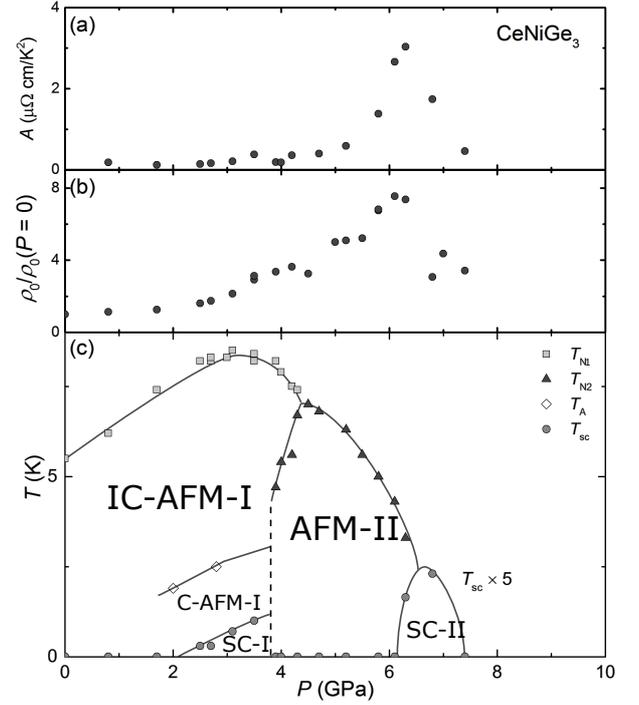}
\end{center}
\caption{Pressure dependence of A coefficient of Fermi liquid term (a), and residual resistivity $\rho_0$ (b).
$P$--$T$ phase diagram of CeNiGe$_{3}$. The squares, triangles, and circles indicate the AFM-I transition temperature $T_{\rm N1}$, the AFM-II transition temperature $T_{\rm N2}$, and SC transition temperature $T_{\rm sc}$ estimated from $\rho(T)$, respectively. The diamonds indicate incommensurate to commensurate transition temperature $T_{\rm A}$ revealed by $^{73}$Ge-NQR measurement\cite{A.Harada_JPSJ_2008}.}
\label{Fig.4}
\end{figure}
%%%%%%%%%%%%%%%%%%%%%%%%%%%%%%%%%%%%%%%%%%%%%%%%%%%%%%%%%%%%%%%%%%%%%%%%%%%

We summarized the $P$--$T$ phase diagram on CeNiGe$_{3}$ in Fig.~\ref{Fig.4}(c).
$T_{\rm N1}$ of the incommensurate AFM-I phase initially increases with increasing pressure, and subsequently decreases above 3~GPa.
From the $^{73}$Ge-nuclear quadrupole resonance (NQR) measurement, it was reported that the incommensurate AFM state transforms into the commensurate AFM state with decreasing temperature above 2~GPa\cite{A.Harada_JPSJ_2008}.
In the single-crystalline sample, the successive transition to AFM-II phase is found above 3.5~GPa, although $T_{\rm N2}$ was not observed in the polycrystalline samples\cite{M.Nakashima_JPCM_2004,H.Kotegawa_JPSJ_2006}.
A magnetoresistance for $H \parallel a$ at 3.9~GPa exhibits a clear hysteresis anomaly below $T_{\rm N2}$, indicating the first-order phase transition from the AFM-II phase to the AFM-I phase\cite{Y.Ikeda_JPSCP_2014}. 
The magnetic structure in AFM II phase is under investigation.
$T_{\rm N2}$ decreases with increasing pressure and becomes zero at around 6.5~GPa.
Furthermore, we revealed two well-separated SC phases.
On the other hand, overlapped two SC phases were observed in the previous report probably due to worse hydrostatic condition and using polycrystalline samples\cite{H.Kotegawa_JPSJ_2006}.
The SC-I phase appears above $\sim$ 2~GPa and simultaneously disappears with the emergence of the AFM-II phase, which means that the SC-I phase coexists with the commensurate AFM-I phase but cannot coexist with the AFM-II phase.
The SC-II phase is located in a narrow pressure range near $P_{\rm c}$.

%%%%%%%%%%%%%%%%%%%%%%%%%%% Figure 5 %%%%%%%%%%%%%%%%%%%%%%%%%%%%%%%%%%%%%
\begin{figure}[!tb]
\vspace*{10pt}
\begin{center}
\includegraphics[width=8cm,clip]{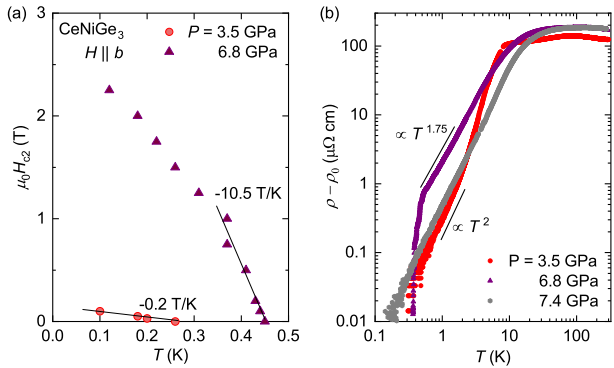}
\end{center}
\caption{(Color online) (a) $\mu_0H_{\rm c2}$ - $T$ phase diagram at 3.5~GPa (SC-I phase) and 6.8~GPa (SC-II phase). (b) Temperature dependence of resistivity subtracted from residual resistivity $\rho_0$ at 3.5 (SC-I phase), 6.8 (SC-II phase), and 7.4~GPa (paramagnetic phase).}
\label{Fig.5}
\end{figure}
%%%%%%%%%%%%%%%%%%%%%%%%%%%%%%%%%%%%%%%%%%%%%%%%%%%%%%%%%%%%%%%%%%%%%%%%%%%

The SC properties are quite different between two phases.
Non-Fermi liquid behavior [$\rho(T) \propto T^{1.75}$] was observed at 6.8~GPa (SC-II phase), which is clearly different from the Fermi liquid behavior [$\rho(T) \propto T^{2}$] at 3.5~GPa (SC-I phase) [See Fig.~\ref{Fig.5} (b)].
Fermi Liquid behavior recovered at 7.4~GPa where the superconductivity disappears.
Note that we applied larger electric current to obtain precise data, and thus, the superconductivity was suppressed when the large electric current was applied.
In addition, the upper critical field $H_{\rm c2}$ of the SC-II phase shows a steep initial slope ($d\mu_0H_{\rm c2}/dT = -10.5$~T/K) compared to that of the SC-I phase ($d\mu_0H_{\rm c2}/dT = -0.2$~T/K) as shown in Fig.~\ref{Fig.5} (a).
It is noted that $\mu_0H_{\rm c2}~\sim~$2.5~T at 6.8~GPa exceeds ordinary Pauli-limitting field $\mu_0H_{\rm P}~\sim~1.84T_{\rm c}~\sim~$0.8~T.
The large initial slope of $\mu_0H_{\rm c2}$ in the SC-II phase is close to $-6$~T/K in CeIn$_3$,  $-5$~T/K in CePd$_2$Si$_2$\cite{F.M.Grosche_JPCM_2001}, and $-20$~T/K in CeRhIn$_5$\cite{G.Knebel_JPSJ_2011}.
The change of the slope is related to the effective mass of electron since the orbital critical field $\mu_0H_{\rm c2}^{\rm orb}$ can be described as $\mu_0H_{\rm c2}^{\rm orb} = 0.693(-d\mu_0H_{\rm c2}/dT)_{T_{\rm sc}}T_{\rm sc} = \Phi_{0}/(2\pi \xi^2) = \pi \Delta_{0}^2 m^{*2}/(2\hslash^4 k_{\rm F}^2)$\cite{R.R.Hake_APL_1967}.
Here, $\Phi_0$ is the quantum fluxoid, $\xi = \hslash^2 k_{\rm F}/\pi \Delta_{0} m^*$ is the SC coherence length, $k_{\rm F}$ is the Fermi wave vector, $\Delta_{0}$ is the SC gap, and $m^*$ is the effective mass of electron.
The enhancement of the effective mass at the critical pressure was also confirmed by the pressure dependence of $A$ coefficient as shown in Fig.~\ref{Fig.4} (a).
For the estimation of $A$ coefficient, we assumed the Fermi liquid behavior [$\rho(T) \propto \rho_0 + A T^2$] at low temperatures.
These results indicate that the heavy quasi-particles form the Cooper pair in the SC-II phase, similar to other heavy fermion superconductors, such as CeIn$_3$ and CePd$_2$Si$_2$.

In contrast with the SC-II phase, the SC-I phase in the AFM-I state has some unusual characteristics.
The SC-I phase is located in the deep inside of the AFM-I state,  which is a rare case on the contrary to superconductivity near a quantum critical point.
Although there are some reports of the coexistence of superconductivity and magnetism in the heavy fermion superconductors, most of those appear in the vicinity of AFM quantum critical point.
To our best knowledge, it has been reported only in some U-based systems such as UNi$_2$Al$_3$ and UPd$_2$Al$_3$\cite{C.Wassilew_PhysicaB_1994,R.Feyerherm_PRL_1994}, and the non-centrosymmetric superconductor CePt$_3$Si\cite{N.Tateiwa_PhysicaB_2006}.
In these systems, the coexistence of magnetic ordering and superconductivity was explained in terms of two rather independent electron subsystems of $f$ character.
One is identified with the itinerant heavy quasi-particle system that is responsible for the superconductivity.
The other one represents more local $f$ electrons and is responsible for the antiferromagnetism\cite{R.Feyerherm_PRL_1994,H.Mukuda_JPSJ_2009-2}.
According to this scenario, the AFM-I phase, which coexists with SC-I phase, originates from localized $f$ electrons, and the AFM-II phase, which cannot coexist with SC-I phase, may have an itinerant character.
If this scenario can be applicable, the change in magnetic character should be a crossover. 
However, the transition from the AFM-I phase to the AFM-II phase was of first-order, which seems to be inconsistent with this scenario.
Another interpretation is the competition between commensurate and incommensurate AFM order.
$^{73}$Ge-NQR results indicate that the SC-I phase only coexists with the commensurate AFM order\cite{A.Harada_JPSJ_2008}.
While the time reversal symmetry and the band degeneracy is preserved in a commensurate AFM order, the band degeneracy is lifted owing to the lack of translational symmetry in an incommensurate AFM order.
Therefore, SC state might be destabilized in an incommensurate AFM order.
The magnetic structure is unclear in the AFM-II phase so far.
In order to understand further detail in the coexistence state, precise microscopic studies such as NMR/NQR and/or neutron scattering experiments are required.

Finally, we discuss the possibility of valence crossover in CeNiGe$_3$.
At ambient pressure, $\rho(T)$ shows the broad maximum at approximately 100~K originating from the splitting of the CEF and the Kondo temperature was estimated as 4.5~K from the specific heat measurements\cite{A.P.Pikul_PRB_2003}.
With applying pressure, the Kondo effect became significant and two peaks in the resistivity were observed.
These two peaks due to the CEF and the Kondo effect merge at around $P_{\rm c}$ = 6.5~GPa as shown in Fig.~\ref{Fig.1}.
In addition, the pressure dependence of residual resistivity $\rho_{0}$ shows a peak at $\sim$ 6.5~GPa as shown in Fig.~\ref{Fig.4} (b).
These features have also been observed in CeCu$_2$Si$_2$ and CeCu$_{2}$Ge$_{2}$ under high pressure, and those have been interpreted as the valence crossover\cite{D.Jaccard_PhysicaB_1999,A.T.Holmes_PRB_2004}.
From the analogy of these results, it is plausible that the valence crossover may occur near $P_{\rm c}$ in CeNiGe$_3$.
However, $T$-linear resistivity, which is the characteristic feature of valence criticality, was not observed in CeNiGe$_3$\cite{S.Watanabe_PRL_2010}.
It is possible that the temperature dependence of resistivity is modified by the combination of magnetic and valence criticality quantum magnetic fluctuations.

%\section{Conclusion}
In conclusion, we have measured the temperature dependence of resistivity with the single-crystalline CeNiGe$_{3}$ sample under high pressure.
In order to establish the characteristic $P$--$T$ phase diagram, the experiments have been performed high-quality single crystal under better hydrostatic conditions.
$T_{\rm N1}$ = 5.5~K at ambient pressure initially increases with increasing pressure and has a maximum at $\sim$ 3~GPa.
Above 2.3~GPa, a clear zero-resistivity was observed and this SC state suddenly disappears at 3.7~GPa simultaneously with the appearance of AFM-II phase.
The AFM-II phase is continuously suppressed with further increasing pressure and disappears at $\sim$ 6.5~GPa.
In the narrow range near the critical pressure, the SC-II phase appears.
The enhancement of effective mass and non-Fermi liquid behavior at $P_{\rm c}$ indicate that the SC-II phase is mediated by AFM fluctuations.
On the other hand, the SC-I phase coexists robustly with the AFM-I phase.
The further understanding of the nontrivial $P$--$T$ phase diagram in CeNiGe$_{3}$ should be helpful for the understanding of the relationship between superconductivity and magnetism in heavy-fermion systems.

%\section*{Acknowledgments}
The authors acknowledge K.~Miyake for fruitful discussions. 
Part of this work was performed at the Advanced Science Research Center, Okayama University. 
This work was partially supported by Grant-in-Aids for Scientific Research (KAKENHI) (Grants No. JP19K14657) and the Program for Advancing Strategic International Networks to Accelerate the Circulation of Talented Researchers from the Japan Society for the Promotion of Science (JSPS).

%\bibliographystyle{apsrev4-1}
%\bibliographystyle{jpsj}
%\bibliographystyle{apsrev4-1_nocomma_etal}
%\bibliography{Ref,NMR}

\end{document}